\documentclass[pra,amsfonts,amssymb,twocolumn,superscriptaddress,showpacs]{revtex4-1}

\usepackage{graphics,graphicx}
\usepackage{amsmath}
\usepackage{times}
\usepackage{amstext}
\usepackage{bbm}

\def\Det{\mbox{Det}}
\def\Tr{\mbox{Tr}}
\providecommand{\openone}{\leavevmode\hbox{\small1\kern-3.8pt\normalsize1}}
\def\rank{\mbox{rank}}
\def\beq{\begin{equation}}
\def\eeq{\end{equation}}
\def\bea{\begin{eqnarray}}
\def\eea{\end{eqnarray}}

\newtheorem{lemma}{Lemma}
\newcommand{\rr}{\mathbbm{R}}
\newcommand{\e}{\mathrm{e}}

\begin{document}

\title{The optimal unitary dilation for bosonic Gaussian channels}

\author{Filippo Caruso}
\affiliation{Institut f\"{u}r Theoretische Physik, Universit\"{a}t Ulm, Albert-Einstein-Allee 11, D-89069 Ulm, Germany}

\author{Jens Eisert}
\affiliation{Institute of Physics and Astronomy, University of Potsdam, 14476 Potsdam, Germany}

\author{Vittorio Giovannetti}
\affiliation{NEST, Scuola Normale Superiore and Istituto Nanoscienze-CNR, Piazza dei Cavalieri 7, I-56126 Pisa, Italy}

\author{Alexander S. Holevo}
\affiliation{Steklov Mathematical Institute, Gubkina 8, 119991 Moscow, Russia}

\begin{abstract}
A generic quantum channel can be represented in terms of a unitary interaction between the information-carrying system and a noisy environment.
Here, the minimal number of quantum Gaussian environmental modes required to provide a unitary
dilation of a multi-mode bosonic Gaussian channel is analyzed both for mixed and pure
environment corresponding to the Stinespring representation. In particular, for the case of pure environment we compute this quantity and present an explicit unitary dilation for arbitrary bosonic Gaussian channel. These results considerably simplify the characterization of these continuous-variable maps and  can be applied to address some open issues concerning the transmission of information encoded in bosonic systems.
\end{abstract}

\pacs{03.67.Hk, 42.50.Dv}

\maketitle

\section{Introduction}\label{s:intro}

Bosonic Gaussian channels (BGCs) are an important special class of
transformations that act on a collection of bosonic modes
preserving the Gaussian character of any Gaussian input quantum
state~\cite{HOLEVOBOOK}. The set of BGCs is singled out from
a theoretical perspective~\cite{HW}, but most significantly also
from the perspective of practical implementations, since it
emerges naturally as the fundamental noise model in several
experimental contexts. In the vast majority of
physical implementations of quantum transmission lines
quantum information is almost invariably sent using photons --- through optical
fibers~\cite{SCHEEL}, in free space~\cite{YS}, or via
superconducting transmission lines~\cite{WALLRAFF}) --- physical
situations for which BGCs provide extraordinarily good models.
What is more, BGCs play a major role in
characterizing the open quantum system dynamics of various setups
which use collective degrees of freedom to store and manipulate
quantum information, including systems from cavity QED,
nano-mechanical harmonic oscillators~\cite{NANO}, or clouds of cold atomic
gases~\cite{SCHORI}.

Unsurprisingly, therefore, the study of Gaussian or quasi-free quantum channels has a long
tradition \cite{HOLEVOBOOK,HW,Old}. Intense recent research has mostly
been focusing on properties of BGCs with respect to their ability to
preserve and transmit quantum information (for a review
see, e.g., Ref.~\cite{EW} and references therein).
Recent contributions include the computation of the quantum
capacity~\cite{QUANTUMCAP}  of a large subset of single mode
BGCs~\cite{WGG}, a characterization in terms of the notion of
degradability --- introduced in Ref.\ \cite{CGH} --- that allows one to
identify a zero-quantum capacity subset of BGCs,
and a necessary and sufficient  condition for BGCs being
entanglement-breaking \cite{ENTBR}.
A general unitary dilation theorem
for BGCs was proven in Ref.~\cite{CEGH}: It shows that each BGC $\Phi$ acting on a system $A$ formed by $n$
input bosonic modes admits a unitary dilation in terms of a  bosonic environment $E$ composed of $\ell \leqslant
2n$ modes, the initial state $\hat{\rho}_E$ of which is Gaussian,
with a Gaussian unitary coupling $\hat{U}_{A,E}$ corresponding to a Hamiltonian
that is quadratic in the canonical coordinates,
  \begin{eqnarray}
  \Phi(\hat{\rho}) = \text{Tr}_E[\hat{U}_{A,E} (\hat{\rho}
\otimes \hat{\rho}_E)\hat{U}_{A,E}^\dagger]\;.
\label{first}\end{eqnarray} Here, $\hat{\rho}$ is the input quantum
state of the system $A$ and $\text{Tr}_E$ denotes the partial trace over
the degrees of freedom associated with $E$.

The fact that the number of environmental modes $\ell$
entering  in the unitary dilation can be bounded from above by $2n$ may be
viewed as the continuous-variable counterpart of the upper bound on
the minimal dilation set by the Stinespring theorem~\cite{STINE} for finite
dimensional quantum channels: It indicates that any quantum channel
can be described by using an
environment which is no more than twice the size of the input
system. An important open question is the characterization
of the minimal value of $\ell^{(\Phi)}$ that is
needed to represent a given BGC, specifically the minimal value $\ell_{\text{pure}}^{(\Phi)}$ in
case of a pure unitary dilation. Similarly to the quantum capacity, this quantity may be used to induce a partial
ordering in the set  of BGCs since, as a general rule of thumb,
one expects that the larger it is,
the noisier and the less efficient in preserving the initial
state will be the associated channel. Furthermore,
an exact estimation of such number will allow one to considerably simplify the degradability analysis
of BGCs by minimizing the number of degrees of freedom of the
corresponding complementary channel.

The main result of this work
is to explicitly identify this minimal value $\ell_{\text{pure}}^{(\Phi)}$ --- so the minimum number
of environmental modes initially in a pure Gaussian state ${{\hat{\rho}}}_E$ in a unitary dilation~(\ref{first})
(related to the Stinespring dilation \cite{footnote}) --- and to construct the corresponding dilation.
To simplify terminology, in this case we speak of Eq.\ (\ref{first}) as of the Stinespring representation.
This is accomplished by first determining a lower bound  for $\ell_{\text{pure}}^{(\Phi)}$ in terms
of the minimum number  $q_{\min}^{(\Phi)}$  of ancillary modes  which are needed to construct a Gaussian purification
of  a (generalized) Choi-Jamiolkowski (CJ)  Gaussian  state of $\Phi$. This is motivated by the fact that
any Gaussian Stinespring representation~(\ref{first}) naturally induces a Gaussian purification of the CJ states of the channel.
Then, we show that this lower bound can be exactly achieved by explicitly constructing a Gaussian Stinespring dilation with $q_{\min}^{(\Phi)}$
modes. In the second part of the paper we finally address the case of unitary dilations~(\ref{first})  in which the environment state $\rho_E$ is not necessarily pure, and provide
an estimation for the minimal $\ell$ which improves the one presented in Ref.~\cite{CEGH}.

The paper is organized as follows:
In Sec.~\ref{s:bos} we
recall some basic definitions and set the notation.
The notion of a generalized CJ state of a BGC and
the lower bound $q_{\min}^{(\Phi)}$ for  $\ell_{\text{pure}}^{(\Phi)}$
are presented
in Sec.\ \ref{s:CJ}. Then, in Sec.\ \ref{s:rec}, we present an explicit recipe to construct such minimal dilation.
The case of dilations involving not necessarily pure environments is finally addressed in Sec.~\ref{s:recmix}, while
conclusions are presented in Sec.~\ref{concl}. This
work includes also some technical appendixes.

\section{bosonic Gaussian channels}\label{s:bos}
Consider a system $A$ composed of $n$ bosonic quantum mechanical modes described by
 the canonical coordinates $\hat{R}:= (\hat{Q}_1, \cdots, \hat{Q}_n; \hat{P}_1,\cdots, \hat{P}_n)$ and by the
{\it Weyl (or displacement) operators}
\begin{equation}
	\hat{V}(z) := \e^{i \hat{R} z},
\end{equation}	
with $z:= (x_1, \cdots, x_n,y_1,\cdots, y_n)^T\in \rr^2$ being a column vector~\cite{HOLEVOBOOK}.
To simplify notation, we choose units in which $\hbar=1$.
A {\it BGC} $\Phi$ acting on $A$ is
completely determined by assigning
 a real vector $v\in \rr^{2n}$  and two  $2n \times 2n$ real matrices
  $Y, X \in  {Gl} ({2n}, \rr)$
satisfying the complete positivity condition
\begin{equation}\label{cptcond}
	Y \geqslant i  \Sigma\,   \qquad \mbox{with} \quad \Sigma :=
	\sigma_{2n}  - X^{T} \sigma_{2n} X  \,,
\end{equation}
where  $\sigma_{2n}$ is the matrix defining the {\it symplectic form}
capturing the canonical commutation relations of
$n$ modes, i.e.,
\begin{eqnarray}
	\sigma_{2n} := {\footnotesize \left[
    	\begin{array}{c|c}
	0 & \openone_n\\ \hline
	-\openone_n & 0
    	\end{array}
    	\right]    \;, }\label{sigma2n}
\end{eqnarray}
with    $\openone_n$ indicating the $n\times n$ identity matrix.
More precisely, the map $\Phi$ is defined   as the linear mapping which, for all $z$ complex, induces the following transformation
\begin{equation}
	\phi(\hat{\rho};z) \mapsto \phi(\Phi(\hat{\rho}); z) :=
	\phi(\hat{\rho}; Xz) \e^{{- \tfrac{1}{4} z^T Y
	z } + i v^T z} \nonumber\;,
\end{equation}
where
\begin{equation}
	\phi(\hat{\rho}; z) := \Tr[ \, \hat{\rho}\, \hat{V}(z) \, ]\;,
\end{equation}	
is the symmetrically ordered {\it characteristic
function} of the state $\hat{\rho}$. A state is called {\it Gaussian} if its characteristic
function is a Gaussian function in phase space \cite{HOLEVOBOOK,Review}. A {\it Gaussian map} is a completely
positive map that maps all unknown Gaussian states onto Gaussian states and a {\it Gaussian unitary}
a unitary generated by a quadratic polynomial in the canonical coordinates, reflected by a {\it symplectic transformation}
from $Sp(2n,\rr)$ on the level of canonical coordinates.

 In the construction of   the Gaussian unitary representations~(\ref{first}) of $\Phi$,
the  vector ${v}$ plays a marginal role since it can be eliminated via a unitary rotation acting on the output state, see, e.g.,
Ref.~\cite{CEGH}.
In contrast, the matrices in Eq.~(\ref{cptcond})
 are of fundamental importance --- in particular, we shall see that the value of $\ell_{\text{pure}}^{(\Phi)}$, and of
 our estimation of  $\ell_{\text{mix}}^{(\Phi)}$,
 depend  upon the
 ranks of $Y$, $\Sigma$, and
$Y-i\Sigma$. It is thus worth anticipating some relevant facts that concern these matrices.
First of all we notice that the
 inequality (\ref{cptcond}) implies the following
relations
\begin{eqnarray} \label{inclu1}
\mbox{ker}[\Sigma] \cap \mbox{ker}[Y-i\Sigma ] &\subseteq &\mbox{ker}
[Y]\subseteq \mbox{ker}[Y-i\Sigma ]\;, \\ \label{inclu2}
\mbox{ker}[Y] &\subseteq &\mbox{ker}[\Sigma]\;,
\end{eqnarray}
where throughout the paper, given a generic (possibly real) $d\times d$ matrix $M$, we
	denote its kernel (null subspace) with $\mbox{ker}[M]:=\left\{ w\in \mathbb{C}^{d}:Mw=0\right\}$~\cite{NOTA}.
The first inclusion in Eq.~(\ref{inclu1}) follows from the
definition, the remaining one and the inclusion of Eq.~(\ref{inclu2}) are  derived from the observation that
$w^{\dag }Yw=0\Rightarrow w^{\dag }(i\Sigma)
w=0\Rightarrow w^{\dag }(Y-i\Sigma )w=0\Rightarrow ( Y-i\Sigma
)w=0\Rightarrow \Sigma w=0$. Putting  these identities together we also find that
\begin{equation} \label{kernelid}
	\mbox{ker}[Y]=\mbox{ker}[\Sigma] \cap \mbox{ker}[Y-i\Sigma ].
	\end{equation}
Other useful properties are the identities
\begin{eqnarray} \label{ideimpo}
2 \; \rank[{Y}-i {\Sigma}] = \rank[{Y}] + \rank[{Y}- {\Sigma} {Y}^{{\ominus1}}\,  {\Sigma}^T ]\;,
\end{eqnarray}
and the inequalities
\begin{equation}
 \rank[Y] \geqslant \rank[{\Sigma}]\geqslant  \rank[{Y}] -  \rank[{Y}- {\Sigma} {Y}^{{\ominus1}}\,  {\Sigma}^T ]\geqslant 0\;,
 \label{ineimpo}
\end{equation}
where $\rank[M]$ stands for   the rank of the matrix $M$ (i.e.
the dimension over the complex field of the complement  to $\mathbb{C}^{d}$ of the matrix  $\mbox{ker}[M]$),
 and ${Y}^{{\ominus1}}$ is the
Moore-Penrose (MP) inverse of $Y$~\cite{nota1}. The explicit proof of these relations is rather
technical and thus we postpone it to Appendix~\ref{appA}. Here we rather point out that the first inequality  of  Eq.~(\ref{ineimpo})
is a consequence of the fact that $\mbox{ker}[Y]$ is included in $\mbox{ker}[\Sigma]$, while the last inequality is an immediate
consequence of the fact that ${\Sigma} {Y}^{{\ominus1}}\,  {\Sigma}^T$ is positive semi-definite.

In Ref.~\cite{CEGH}, an upper bound for  $\ell_{\text{pure}}^{(\Phi)}$
was set by showing that one can construct a Stinespring dilation
of $\Phi$ that involves
$\ell = 2n - r'/2$
environmental modes with
\begin{equation}
	r^\prime :=\rank[Y]- \mbox{rank}[ Y - \Sigma
	Y^{\ominus1} \Sigma^T].
\end{equation}	
In what follows we  will strengthen this result by showing that the minimum number of modes necessary to build
a Gaussian  Stinespring unitary dilation for $\Phi$ is given by  
\begin{eqnarray} \label{boundmin}
\ell_{\text{
pure}}^{(\Phi)} = \rank[Y] -r^\prime/2 = \rank[ Y - i \Sigma] \;,
\end{eqnarray}
where
we used Eq.~(\ref{ideimpo}) when formulating the second identity.
Since $Y$ is a $2n\times 2n$ matrix, we have
$ 2n -k \geqslant 0$, and so the optimal bound we prove here leads to a
significant improvement compared to the results of Ref.~\cite{CEGH}.  In particular, for those BGCs which represent unitary transformations
of the $n$ input modes (i.e., $Y=0$ and $X\in Sp(2n,\rr)$
symplectic~\cite{HOLEVOBOOK}) the optimal bound~(\ref{boundmin}) yields
$\ell_{\text{ pure}}^{(\Phi)}= 0$ --- no environment is
required to construct the dilation --- while Ref.~\cite{CEGH} had this value equal to $2n$.
To prove Eq.~(\ref{boundmin}) we shall first show that the quantity $\rank[ Y - i \Sigma]$ provides a lower bound for  $\ell_{\text{
pure}}^{(\Phi)}$ (see Sec.~\ref{s:CJ}) and then construct an explicit
Stinespring dilation~(\ref{first}) for $\Phi$  which attains such bound (see Sec.~\ref{s:rec}).

\section{Lower bound on $\ell_{\text{pure}}^{(\Phi)}$ via Generalized CJ-states of  BGCs} \label{s:CJ}

In this section we review the notion of generalized  Choi-Jamiolkowski (CJ) state for a multi-mode BGC (see also Ref.~\cite{notesholevo}
and compare Refs.\ \cite{Fiurasek}), and use it to
show that the term on the rhs.\  of Eq.~(\ref{boundmin}) provides a lower bound for  $\ell_{\text{pure}}^{(\Phi)}$.
Consider a state vector $|\Psi_{\hat{\Lambda}} \rangle_{A,B}$ providing a purification of a quantum
state $\hat{\Lambda} = \sum_{j=1}^\infty \lambda_j (| j\rangle\langle j|)_A$ of the
system labeled $A$ which has full rank  (e.g., a Gibbs state of $n$ modes).
That is to say,
\begin{eqnarray}
|\Psi_{\hat{\Lambda}} \rangle_{A,B} &=& \sum_{j=0}^\infty\sqrt{\lambda_j} | j\rangle_A \otimes |j \rangle_B
\nonumber \\
&=& ({\hat{\Lambda}}^{{1}/{2}} \otimes \openone) \sum_{j=0}^\infty | j\rangle_A \otimes |j \rangle_B \; , \nonumber
\end{eqnarray}
with $A$ indicating the input space of the channel $\Phi$, $B$ being an ancillary system  isomorphic to $A$,
and $\{ |j\rangle : j=0,\cdots ,\infty\}$ denoting an orthonormal complete basis.
A generalized CJ state of the channel $\Phi$ is now obtained as
\begin{eqnarray}\label{fggg}
	\hat{\rho}_{A,B}(\Phi)=\left(\Phi \otimes {\cal I} \right)( | \Psi_{{\hat{\Lambda}}} \rangle
	\langle \Psi_{{\hat{\Lambda}}} |)_{A,B}\;,
\end{eqnarray}
with ${\cal I}$ being the identity map.
The state $\hat{\rho}_{A,B}(\Phi)$ provides a complete  representation of the channel via the inversion formula,
\begin{eqnarray}\label{eq}
	\Phi(\hat{\rho}) = \mbox{Tr}_{B} [ (\openone_A\otimes {\hat{\Lambda}}_B^{-1/2} {\hat{\rho}}_B^T  {\hat{\Lambda}}_B^{-1/2} ) \hat{\rho}_{A,B}(\Phi)]\;,
\end{eqnarray}
where ${\hat{\rho}}_B$ and ${\hat{\Lambda}}_B$ are copies of the states ${\hat{\rho}}$ and ${\hat{\Lambda}}$ on $B$,
respectively,
while ${\hat{\rho}}_B^T$ is its transpose with respect to the orthonormal  basis introduced above. We will suppress an index
labeling both the chosen basis and the reference state.

For finite-dimensional system $\hat{\rho}_{A,B}(\Phi)$ provides a standard CJ state representation
when ${\hat{\Lambda}}$ is taken to be the maximally mixed state (compare Refs.\
\cite{Fiurasek}). In the infinite-dimensional case
such limit in general is well defined only in the context of positive forms,
see Ref.~\cite{notesholevo}.
However, Eq.~(\ref{eq}) shows  that we do not need to approach such a limit in order to build a proper
representation of the channel: It is defined for any state diagonal in the distinguished basis of full rank.
Furthermore, it is easy to verify that it is always possible to work with  CJ states $\hat{\rho}_{A,B}(\Phi)$ which
are Gaussian: To do so simply take $(|\Psi_{\hat{\Lambda}} \rangle\langle\Psi_{\hat{\Lambda}}|)_{A,B}$ to be Gaussian and use the fact that
the Gaussian map $\Phi \otimes {\cal I}$ maps Gaussian states into Gaussian states.
In the following we choose to take such Gaussian reference states. In particular, we will
assume $(|\Psi_{\hat{\Lambda}} \rangle\langle\Psi_{\hat{\Lambda}}|)_{A,B}$ to be a Gaussian purification of
a multi-mode Gibbs (thermal) state of quantum mechanical oscillators.

An important observation concerning the generalized CJ representation is that,
given a Stinespring representation of $\Phi$ involving an environmental system $E$, one can construct a purification of $\hat{\rho}_{A,B}(\Phi)$ that uses $E$ as ancillary system.
Indeed, assuming that   $\hat{U}_{A,E}$ and $(|0\rangle\langle 0|)_E$ give rise to a Stinespring representation for $\Phi$, we have that
the pure state with state vector
\begin{eqnarray} \label{chieq}
|\chi\rangle_{A,B,E} = \hat{U}_{A,E} |\Psi_{\hat{\Lambda}} \rangle_{A,B}\otimes |0\rangle_E
 \end{eqnarray}
 is a purification of $\hat{\rho}_{A,B}(\Phi)$. Furthermore, if $\hat{\rho}_{A,B}(\Phi)$ is Gaussian and
 $E$ represents a collection of $\ell$ environmental bosonic modes with $|0\rangle_E$ being a Gaussian state vector
 and $U_{A,E}$ being a Gaussian unitary,
 it follows that also $|\chi\rangle_{A,B,E}$ will define a Gaussian purification of $\hat{\rho}_{A,B}(\Phi)$. Putting
 these facts together it follows that
 a lower bound for the minimal number $\ell_{\text{pure}}^{(\Phi)}$ of environmental modes that are needed to build a Gaussian Stinespring representation of $\Phi$ is provided by the
 minimal number $q_{\text{min}}^{(\Phi)}$ of Gaussian ancillary modes that are required
 to purify a generalized  Gaussian CJ state $\hat{\rho}_{A,B}(\Phi)$ of $\Phi$, i.e., we have that
 \begin{eqnarray}
 \ell_{\text{pure}}^{(\Phi)}  \geqslant q_{\text{min}}^{(\Phi)}
 \label{ineqqq}\;.
 \end{eqnarray}
To compute $q_{\text{min}}^{(\Phi)}$ we first make a specific choice for $|\Psi_{\hat{\Lambda}} \rangle_{A,B}$. In particular,
since $A$  is composed by $n$ bosonic modes, we can take $|\Psi_{\hat{\Lambda}} \rangle_{A,B}$ to be a
product  of $n$ identical two-mode state vectors
of the form
\begin{eqnarray}
	|\Psi_{\hat{\Lambda}} \rangle_{A,B} = \bigotimes_{i=1}^n |\psi\rangle_{A_i,B_i}
\end{eqnarray}
where $|\psi\rangle_{A_iB_i}$ reflects
a purification of a Gibbs state of the $i$-th mode $A_i$ of $A$ built by coupling it with the corresponding ancillary system $B_i$:
This is nothing but what is usually referred to as a {\it two-mode squeezed state} \cite{Review}.
The resulting state is of course Gaussian and it is fully characterized by its covariance matrix. To express it in a compact form
note that the kernel of the natural symplectic form for the $2n$ modes of
$A,B$ is given by
\begin{equation}\label{NS}
	\sigma_{A,B}: = {\footnotesize \left[
   	 \begin{array}{c|c}
	\sigma_{2n} & 0 \\ \hline
    	0 & \sigma_{2n}
    	\end{array}
    	\right] },
\end{equation}	
where the upper-left and lower-right block matrices represent the symplectic forms of the $n$ modes of
$A$ and $B$, respectively, defined as in Eq.~(\ref{sigma2n}).
With this choice the covariance matrix $\gamma$ of $(|\Psi_{\hat{\Lambda}} \rangle_{A,B}\langle \Psi_{\hat{\Lambda}}|)$
is given by the following $ {Gl} ({4n}, \rr)$ matrix,
\begin{eqnarray}
	\gamma =
 	{\footnotesize \left[
    	\begin{array}{c|c}
	\alpha & \delta  \\ \hline
  	\delta^T & \beta
    	\end{array}
    	\right]     }\;,
\end{eqnarray}
where $\alpha$, $\beta \in {Gl} ({2n},
\rr)$ are the covariance matrices of the $A$ and $B$ modes, respectively, with $\delta$, $\delta^T  \in {Gl} ({2n},
\rr)$ being the cross-correlation terms. Explicitly, they are given by
\begin{eqnarray}
  	{\footnotesize
   	\alpha  =
     	\left[
          \begin{array}{c|c}
    \theta\openone_{n}  & 0 \\ \hline
    0 &   \theta\openone_{n}
    \end{array} \right] = \beta\;,}
    \; \; \;
     \delta =
     \left[
          \begin{array}{c|c}
   0 & f(\theta)  \openone_{n}  \\ \hline
     f(\theta)  \openone_{n} &  0
    \end{array} \right] = \delta^T \; , \nonumber
    \end{eqnarray}
    with $\theta > 1$ and
    \begin{equation}
    f(\theta) := - ({\theta^2 -1})^{1/2}.
    \end{equation}
    The parameter
    $\theta$ determines the {\em temperature} of the Gibbs states we used to build the vector $|\Psi_{\hat{\Lambda}} \rangle_{A,B}$, or
    equivalently, the {\it two-mode squeezing parameter} of the purification.
    In particular, the case $\theta=1$ corresponds to the limit in which all the modes of $A$ and $B$ are prepared into the vacuum state:
    In this case the state ${\hat{\Lambda}}$ no longer has
    maximum support and thus does not provide a proper starting point to build a CJ state. For $\theta\rightarrow \infty$, in contrast,
    the state  $|\Psi_{\hat{\Lambda}} \rangle_{A,B}$ approaches a purification of a maximally mixed state for the modes
    (for details see Ref.~\cite{notesholevo}). Equivalently, it corresponds to the limit of large squeezing in the two-mode squeezed
    state of the purification.
    Notice also that by construction, for all values of $\theta \geqslant 1$, $\gamma$ satisfies the
    condition $\gamma \geqslant i \sigma_{A,B}$, as it indeed represents a physical pure state.

The generalized CJ state $\hat{\rho}_{A,B}(\Phi)$ for a Gaussian channel characterized by matrices $Y$ and $X$ as in Eq.~(\ref{cptcond}) is now computed as in Eq.~(\ref{fggg}).
The resulting state is still Gaussian and has the covariance matrix $\gamma' \in {Gl} ({4n},
\rr)$ given by
\begin{eqnarray}
\gamma' =
 {\footnotesize \left[
    \begin{array}{c|c}
X^T \alpha X + Y & X^T \delta  \\ \hline
  \delta^T X & \beta
    \end{array}
    \right]     }
={\footnotesize \left[
    \begin{array}{c|c}
\theta X^T  X + Y &  f(\theta) X^T \sigma_x \\ \hline
  f(\theta) \sigma_x  X & \theta \openone_{2n}
    \end{array}
    \right]  \;,  } \nonumber
\end{eqnarray}
where
\begin{equation}
\sigma_x:={\footnotesize \left[
          \begin{array}{c|c}
    0 &\openone_{n}   \\ \hline
     \openone_{n} &0
    \end{array} \right]}.
 \end{equation}
In general it will be a mixed state and we are interested in
the minimum number $q_{\min}^{(\Phi)}$ of
ancillary modes $q$ that is needed to construct a Gaussian purification of it.
As discussed in  Appendix~\ref{ap:B}, this is given by the quantity
\begin{eqnarray}
q_{\min}^{(\Phi)}&=& \rank[\gamma' - i \sigma_{A,B}] - 2n\nonumber \\
&=&  2n - \dim\mbox{ker}[\gamma' - i \sigma_{A,B}]
\;, \label{ffee}
\end{eqnarray}
(note that in this case $\gamma',\sigma_{A,B}\in Gl(4n\times 4n,\rr)$).
  In what follows we will compute this quantity, showing that it
    coincides with the right hand side of Eq.~(\ref{boundmin}).
      To do so, we first notice that the dimension of the kernel of $\gamma' - i \sigma_{A,B}$ can be expressed as
   \begin{equation}
\dim{\mbox{ker}}[\gamma' - i \sigma_{A,B}]
   = \dim{\mbox{ker}}
 {\footnotesize \left[
    \begin{array}{c|c}
\theta X^T  X + Y -i \sigma& f(\theta) X^T  \\ \hline
  f(\theta)   X & \theta \openone_{2n} + i \sigma
    \end{array}
    \right]     }\;,
\label{ffee1}
\end{equation}
where the
second identity was obtained
 by rotating $\gamma' - i \sigma_{A,B}$ with the transformation
\begin{equation}
	T:= {\footnotesize \left[
          \begin{array}{c|c}
    \openone_{2n} & 0    \\ \hline
   0  &  \sigma_x
    \end{array} \right]}.
 \end{equation}
As for any positive
semi-definite matrix $M$, the
kernel in Eq.~(\ref{ffee1})  can be computed as the set of vectors $w\in \mathbb{C}^{d}$ which satisfy the condition $w^{\dag}Mw =0$~\cite{NOTA}.
Writing $w=(w_{A},w_{B})$, we arrive at the condition
\begin{eqnarray}\nonumber
&&
\theta \left( w_{A}^{\ast
}X^{T}Xw_{A}-w_{A}^{\ast }X^{T}w_{B}-w_{B}^{\ast }Xw_{A}+w_{B}^{\ast
}w_{B}\right)\\
&&\qquad\qquad  +\; w_{A}^{\ast }\left( Y-i\sigma \right) w_{A}+w_{B}^{\ast }i\sigma
w_{B}\nonumber \\
&&\qquad \qquad +\; O(1/\theta )\left( w_{A}^{\ast }X^{T}w_{B}+w_{B}^{\ast }Xw_{A}\right)
=0,  \label{ker1}
\end{eqnarray}
where in the first and second  line we have collected all terms which are linear and constant in $\theta$, respectively.
For $\theta>1$ sufficiently large this requires the following conditions,
\begin{eqnarray}
	w_{A}^{\ast }X^{T}Xw_{A}-w_{A}^{\ast }X^{T}w_{B}&-&w_{B}^{\ast
	}Xw_{A}
	\nonumber\\
	+w_{B}^{\ast }w_{B} &=&0\;,  \\
	w_{A}^{\ast }\left( Y-i\sigma \right) w_{A}+w_{B}^{\ast }i\sigma w_{B} &=&0\;.
\end{eqnarray}
The first equation means $Xw_{A}=w_{B},$ whereas the second reads $
w_{A}^{\ast }\left( Y-i\sigma \right) w_{A}+w_{A}^{\ast }iX^{T}\sigma
Xw_{A}=0,$ that is
\begin{equation}
w_{A}^{\ast }[Y-i\Sigma ]w_{A}=0.  \label{ker2}
\end{equation}
There is one-to-one correspondence between solutions $w_{A}$ of
Eq.~(\ref{ker2}) and $w=(w_{A},Xw_{A})$ of Eq.~(\ref{ker1}), hence
\begin{equation*}
\dim \mbox{ker}[\gamma ^{\prime }-i\sigma _{A,B}]=\dim \mbox{ker}[Y-i\Sigma ].
\end{equation*}
Replacing this into Eq.~(\ref{ffee})  we finally get
\begin{eqnarray}
q_{\min}^{(\Phi)}
=2n -\dim \mbox{ker}[Y-i\Sigma ] =
\rank[ Y  - i \Sigma]
\;, \label{ffeeffaf}
\end{eqnarray}
where in the last identity we used the fact that $Y - i \Sigma$ is a $2n\times 2n$ matrix.

\section{Optimal bound and explicit construction}\label{s:rec}
In this section  we explicitly construct a Gaussian unitary dilation with $q_{\min}^{(\Phi)}=\rank[ Y  - i \Sigma]$
environmental modes. In this way, we demonstrate
that the lower bound derived in the previous section is tight, concluding the derivation of  Eq.~(\ref{boundmin}).
To do so, let us assume that the number of modes which define the state $\hat{\rho}_E$ in Eq.~(\ref{first}) are $q_{\min}^{(\Phi)}$.
Without loss of generality, we write the
kernel of the form corresponding to the commutation relations of our $n + q_{\min}^{(\Phi)}$ modes in
block structure
\begin{eqnarray}
{\sigma} :=  \sigma_{2n} \oplus \sigma^{E}_{2 q_{\min}^{(\Phi)}} =
     { \left[
    \begin{array}{c|c}
     \sigma_{2n}&{0} \\ \hline
    {0} & \sigma^{E}_{2 q_{\min}^{(\Phi)}}    \end{array}
    \right]}\;,
    \end{eqnarray}
where $\sigma_{2n}$ and $\sigma^{E}_{2q_{\min}^{(\Phi)}}$  are $2n\times2n$ and
$2q_{\min}^{(\Phi)} \times 2q_{\min}^{(\Phi)}$ matrices associated with the system and
environment, respectively. While $\sigma_{2n}$ is defined as in
Eq.~(\ref{cptcond}), for $\sigma^{E}_{2 q_{\min}^{(\Phi)}}$ we do not make any
assumption at this point, leaving open the possibility of defining
it later on. Accordingly, the Gaussian unitary
$\hat{U}_{A,E}$ of Eq.~(\ref{first}) will be determined by a
symplectic  matrix $S\in Sp(2(n+q_{\min}^{(\Phi)}),\rr)$ of block form
 \begin{equation}
 S := {\footnotesize
     \left[
    \begin{array}{c|c}
   s_1 & s_2 \\\hline
    s_3 & s_4    \end{array}
    \right]}
 \end{equation}
satisfying the condition $S  {\sigma} S^T = {\sigma}$. In the
above expressions, $s_1$ and  $s_4$ are  $2n\times2n$ and $2 q_{\min}^{(\Phi)}
\times 2 q_{\min}^{(\Phi)}$ real square matrices, while $s_2$ and $s_3^T$ are $2 n
\times 2 q_{\min}^{(\Phi)}$ real  rectangular matrices. As noticed in
Ref.~\cite{CEGH}, the possibility of realizing the unitary
dilation~(\ref{first}) can now be proven by simply taking
\begin{equation}
	s_1 = X^T
\end{equation}	
and finding   $s_2$ and a $q_{\min}^{(\Phi)}$-mode covariance
matrix~\cite{HOLEVOBOOK} $\gamma_E \geqslant i \sigma_{2 q_{\min}^{(\Phi)}}^E$
satisfying the conditions
 \begin{eqnarray}
    s_2 \, \sigma^{E}_{2 q_{\min}^{(\Phi)}} \, s_2^T = \Sigma \,, \qquad
    s_2 \, \gamma_E \, s_2^T = Y \label{probb1} \,,
\end{eqnarray}
with $\gamma_E$ being the covariance matrix of the
Gaussian state $\hat{\rho}_E$ of Eq.~(\ref{first}).

First, let  us consider the case in which $q_{\min}^{(\Phi)}$ is an even
number. To identify valid $s_2$ and $\gamma_E$ which solve
Eq.~(\ref{probb1}), it is useful to
transform $Y$ and $\Sigma$ as in Eq.\ (\ref{y1111}) and (\ref{sigma2222}) of Appendix \ref{appA} (take $A=T$, $B=\Sigma$,
$m=2n$, $a=k$, and $b=r=\rank[\Sigma]$).
Actually, applying an extra orthogonal matrix, $Y'$ is still like in (\ref{y1111}), while $\Sigma'$ can be written as
\begin{eqnarray}  \label{Mequ1}
{\footnotesize \Sigma^\prime :=C \Sigma C^{T}=
     \left[
    \begin{array}{c|c|c}
    {0} &  \begin{array}{c|c}
    \mu  &  0 \\ \hline
    0 &  {0}
    \end{array} & 0
     \\ \hline
     \begin{array}{c|c}
    - \mu& 0 \\ \hline
    0  &{0}
    \end{array}
     & {0} & 0\\ \hline
     0 & 0 & 0
       \end{array}
    \right] \begin{array}{l}
    \} \, r/2 \\
    \} \, (q_{\min}^{(\Phi)}-r)/2\\
     \} \, r/2 \\
    \} \, (q_{\min}^{(\Phi)}-r)/2 \\
    \} \, 2n-q_{\min}^{(\Phi)}, \\
    \end{array}}
    \end{eqnarray}
where $C \in {Gl} ({2n}, \rr)$ and $\mu= \mbox{diag}(\mu_1,\cdots, \mu_{r/2})$ is the
$r/2\times r/2$  diagonal matrix formed by the strictly positive
eigenvalues of $|\Sigma^\prime|$ (satisfying
$\openone_{r/2}\geqslant \mu$ as in Appendix \ref{appA}). Introducing then
$s_2^\prime := C\,  s_2$ the conditions of Eqs.\ (\ref{probb1}) can be
equivalently written as
\begin{eqnarray}
    s_2^\prime \, \sigma_{2 q_{\min}^{(\Phi)} }^E
    \,(s_2^\prime)^T = \Sigma^\prime\,, \quad  \quad
    s_2^\prime \,  \gamma_E  \, (s_2^\prime)^T =
    Y^\prime\,.\label{prob1}
\end{eqnarray}
The explicit expressions for corresponding  $\gamma_E$
and $s_2$ are obtained in the following way. We take the
environmental symplectic form to be
\begin{equation}
	\sigma^{E}_{2 q_{\min}^{(\Phi)}} = \sigma_{k}
	\oplus \sigma_{k -r^\prime}
\end{equation}
where we have set $k:=\rank[Y]$. A unitary dilation with
$q_{\min}^{(\Phi)} = k -r^\prime/2$ environmental modes in a pure
state is obtained by choosing the $2n\times 2q_{\min}^{(\Phi)}$ rectangular matrix $s_2^\prime$ as
\begin{eqnarray}
 s^\prime_2 = {\footnotesize \left[
    \begin{array}{c|c}
    \tilde{K}^{-1} & A \\ \hline
    0 & 0
    \end{array}
    \right]
    \;,}
 \end{eqnarray}
with $\tilde{K}$ being the $k \times k$ symmetric matrix
defined by
 \begin{eqnarray}
 {\footnotesize \tilde{K} :=
\left[
    \begin{array}{c|c}
    \begin{array}{c|r}
    \mu^{-1/2}  & \, 0 \\ \hline
    0 & \openone_{(k-r)/2}
    \end{array} &  0
     \\ \hline
    0
     &  \begin{array}{c|r}
    \mu^{-1/2} & 0 \\ \hline
    0 & \openone_{(k-r)/2}
    \end{array}    \end{array}
    \right]
     \label{ktilde}}
    \end{eqnarray}
and $A$ being a
rectangular matrix $k\times (k-r^\prime)$ of the form
 \begin{eqnarray}
 {\footnotesize
A := \left[
    \begin{array}{c|c}
    {0} &  \begin{array}{c|c}
      0 & 0 \\ \hline
    0 &  0
        \\ \hline
        0 & \openone_{(k-r)/2}
    \end{array}
        \\ \hline
\begin{array}{c|c}
      0 & 0 \\ \hline
    0 &  0
        \\ \hline
        0 & \openone_{(k-r)/2}
    \end{array}
    & {0}
    \end{array}
    \right] \begin{array}{l}
    \} \, r^\prime/2 \\
    \} \, (r-r^\prime)/2 \\
    \} \, k/2-r/2 \\
    \} \, r^\prime/2 \\
    \} \, (r-r^\prime)/2 \\
    \} \, k/2-r/2 .\\
    \end{array}}
    \label{defAnew}
    \end{eqnarray}
By direct substitution one can easily verify that the first
condition of Eq.~(\ref{prob1}) is indeed satisfied. Vice versa,
expressing the $(2k -r')\times (2k-r')$ covariance matrix of $\hat{\rho}_E$ as
\begin{equation}
\gamma_E= \left[\footnotesize{
    \begin{array}{c|c}
    \alpha & \delta \\ \hline
    \delta^T & \beta
     \end{array}}
    \right]\;,
\end{equation}
the second condition of Eq.~(\ref{prob1}) yields  the following
equation
 \begin{eqnarray} \label{NNequation}
\alpha + A\,  \delta^T + \delta \, A^T + A \, \beta \, A^T =
\tilde{K}^{2} \,.
 \end{eqnarray}
A solution can be easily derived by taking the $k\times k$ block $\alpha$ as
\begin{eqnarray}
  {\footnotesize
   \alpha =
     \left[
    \begin{array}{c|c}
        \begin{array}{c|c}
    \mu^{-1} & 0 \\ \hline
    0 &  \tfrac{5}{4} \openone_{(k-r)/2}
    \end{array}
& 0 \\ \hline
    0 &          \begin{array}{c|c}
    \mu^{-1} & 0 \\ \hline
    0 &  \tfrac{5}{4} \openone_{(k-r)/2} \\
    \end{array}
    \end{array}
    \right]}\label{alphadef}
    \end{eqnarray}
while $\beta$ and $\delta$ are, respectively,  $(k -r^\prime)
\times (k-r^\prime)$ and $k\times (k-r')$ real
 matrices defined as follows:
      \begin{eqnarray}
  {\footnotesize \beta:=
     \left[
    \begin{array}{c|c}
        \begin{array}{c|c}
   \mu_o^{-1} & 0 \\ \hline
    0 &  \tfrac{5}{4} \openone_{(k-r)/2}
    \end{array}
& 0 \\ \hline
    0 &          \begin{array}{c|c}
  \mu_o^{-1} & 0 \\ \hline
    0 &   \tfrac{5}{4}\openone_{(k-r)/2}
    \end{array}
    \end{array}
    \right] ,
}
    \end{eqnarray}
 \begin{eqnarray}
  {\footnotesize  \delta := \left[
    \begin{array}{c|c}
        0
& \begin{array}{c|c} 0 & 0 \\ \hline
 f(\mu_o^{-1})
 & 0 \\ \hline
    0 & - \tfrac{3}{4}  \openone_{(k-r)/2}
    \end{array} \\ \hline
     \begin{array}{c|c}
     0&0 \\ \hline
    f(\mu_o^{-1}) & 0 \\ \hline
    0 &  - \tfrac{3}{4}  \openone_{(k-r)/2}
    \end{array}  &  0
    \end{array}
    \right],
  }  \end{eqnarray}
with $\mu_o$ is the $(r-r^\prime)/2\times (r-r^\prime)/2$ diagonal
matrix formed by the elements of $\mu$ which are strictly smaller
than $1$, and with $f(\theta)$ defined as in Sec.\ \ref{s:CJ}. Notice that the parameter $r'$ (defined above)
corresponds also to the number of eigenvalues having modulus $1$ of
the matrix $\Sigma^\prime$, i.e.,
\begin{equation}
	r'= 2n  -
	\mbox{rank}[ \openone_{2n} - \Sigma^\prime (\Sigma^{\prime})^T],
\end{equation}
as can be easily shown by using Eq.\ (\ref{b18}) with $A=Y$ and $B=\Sigma$.
With the choice we made on the commutation matrix $\sigma^{E}_{2 q_{\min}^{(\Phi)}}$,
the matrix $\alpha$ is a $k\times k$ covariance
matrix for a set of independent $k/2$ bosonic modes, the matrix $\beta$
is a $(k -r^\prime)
\times (k-r^\prime)$ covariance
matrix for a set of independent $(k-r^\prime)/2$ modes, and
the matrices $\delta$ and $\delta^T$ represent cross-correlation
terms among such sets. For all diagonal matrices $\mu$ compatible
with the constraint
\begin{equation}
	\openone_{r/2} \geqslant \mu \;,
\end{equation}	
the solution
$\gamma_E$ satisfies also the uncertainty relation  $\gamma_E
\geqslant i \sigma^{E}_{2 q_{\min}^{(\Phi)}}$. Furthermore, since it has
\begin{equation}
	\Det[
	\gamma_E] =1,
\end{equation}	
this is also a minimal uncertainty state, i.e., a
pure Gaussian state of $q_{\min}^{(\Phi)}$ modes~\cite{HOLEVOBOOK}.
By a close inspection of the covariance matrix $\gamma_E$ derived
above, one realizes that it is composed of three
independent pieces. The first one describes a collection of
$r^\prime/2$ vacuum states. The second one, in turn, describes
$(r-r^\prime)/2$ thermal states characterized by the matrices
$\mu_o^{-1}$ which have been purified by adding further
$(r-r^\prime)/2$ modes. The third one, finally,  reflects a
collection of  $k -r$ modes prepared in a pure state formed by
$k/2-r/2$ independent pairs of modes which are entangled. Let us point out again that this covariance matrix is indeed formed
by $q_{\min}^{(\Phi)}$ modes.
The whole derivation can be trivially  extended for $q_{\min}^{(\Phi)}$ odd, by adding to the previous covariance matrix a single
mode in the vacuum state.
\section{Dilations with mixed environments}\label{s:recmix}
In Ref. \cite{CEGH} it was shown that for arbitrary (not necessarily Stinespring)
dilations one can consider an environment of only  $\ell = 2n - r/2$ modes --- observe that $r$ is larger than the quantity $r'$ introduced in Sec.~\ref{s:bos} because of Eq.~(\ref{ineimpo}).
Here, we will strengthen this bound by showing that it is possible
to construct a unitary dilation using just
 \begin{eqnarray} \label{boundminmix}
\ell_{\text{mix}}^{(\Phi)} = k-r/2 = \rank[Y]-\rank[\Sigma]/2 \;,
 \end{eqnarray}
environmental modes which are prepared in a Gaussian, but
not necessarily pure, state. Note that  the term on the rhs.\   is nonnegative due to the
first of the inequalities in Eq.~(\ref{ineimpo}), and that it is explicitly smaller than the
one provided in Ref.~\cite{CEGH} due to the fact that $Y$ is a $2n\times 2n$ matrix. It is worth stressing however that differently from the pure dilation case,  we are not
able to determine whether Eq.~(\ref{boundminmix}) is indeed the optimal bound (we believe it is).

For the sake of simplicity, again
we will treat explicitly only the case of $k$ even (the analysis
however can be easily extended to the odd case).
Because of the structure of $A$ given in Eq.~(\ref{defAnew}), the $(k-r)$ environmental modes prepared in a pure state (see the end of Sec.\ \ref{s:rec})
enter explicitly in the identity in Eq.~(\ref{NNequation}):
consequently, if we wish to satisfy such relation, we cannot remove
any of these modes without changing $A$. Vice versa we can drop
some of the auxiliary modes which were introduced only
for purifying the environmental state. Since they are $(r-r')/2$,
we can reduce the number of modes from  $\ell_{\text{pure}}^{(\Phi)}$ to
\begin{equation}
	\ell_{\text{mix}}^{(\Phi)}= \ell_{\text{pure}}^{(\Phi)} - (r-r')/2 =  k -r/2.
\end{equation}	
To see this explicitly, take
\begin{equation}
	\sigma_{2 \ell_{\text{mix}}^{(\Phi)}}^E = \sigma_k \oplus \sigma_{k -r}.
\end{equation}
The matrix
$s_2^\prime$ can be still expressed as above
but with $A$ being a rectangular matrix $k\times (k-r)$ of the
form
 \begin{eqnarray}
 {\footnotesize
  A := \left[
    \begin{array}{c|c}
    {0} & \begin{array}{c}
  0 \\  \hline
 \openone_{(k-r)/2}
    \end{array}
        \\ \hline
     \begin{array}{c}
    0 \\ \hline
     \openone_{(k-r)/2}
    \end{array}
    & {0}
    \end{array}
    \right] } \;.
    \end{eqnarray}
Similarly, $\beta$ and $\delta$ entering in the definition of
$\gamma_E$ become, respectively, the following $(k -r)\times
(k-r)$ and
 $k \times (k-r)$ real matrices:
     \begin{eqnarray}
     {\footnotesize
    \beta :=
     \left[
    \begin{array}{c|c}
        \tfrac{5}{4} \openone_{(k-r)/2}
& 0 \\ \hline
    0 &          \tfrac{5}{4} \openone_{(k-r)/2}     \end{array}
    \right]}
    \label{Nbeta}
    \end{eqnarray}
 and
  \begin{eqnarray}
    \delta :=
    {\footnotesize \left[
    \begin{array}{c|c}
      0 & 0 \\ \hline
    0 &  -\tfrac{3}{4}  \openone_{(k-r)/2}
        \\ \hline
        0 & 0 \\ \hline
    -\tfrac{3}{4}  \openone_{(k-r)/2} & 0
    \end{array}
    \right] \;.}
    \end{eqnarray}
This covariance matrix now consists of two independent parts: the
first one describes a collection of $r/2$ thermal states described
by the matrices $\mu^{-1}$. The second one reflects a
collection of $k -r$ modes prepared in a pure state formed by
$k/2-r/2$ independent couples of modes which are entangled.
\section{Conclusions}
\label{concl}

We have analytically computed the minimum number of environmental modes necessary for a Gaussian unitary
dilation of a generic multi-mode bosonic Gaussian channel. Moreover, we have also explicitly demonstrated how to construct such a
Gaussian dilation in terms of the covariance matrix of the noisy environment and the symplectic transformation associated to the unitary system-environment interaction. These results may allow one to introduce a classification of the bosonic Gaussian channels in terms of the corresponding noise induced by these maps, which is somehow related to the minimum number of environemntal modes to represent such channels. Moreover, constructing a dilation with a minimal number of auxiliary modes may be useful to minimize the size of the corresponding complementary channel and then to simplify the degradability analysis, which is extremely useful in the calculation of the quantum capacity of these continuous-variable quantum maps.

\section*{Acknowledgements}

This work was in part supported by the FIRB IDEAS project RBID08B3FM.
F.\ C.\ was supported also by a Marie Curie Intra European Fellowship within the 7th
European Community Framework Programme. A.\ H.\ acknowledges partial support of the RFBR grant 09-01-00424 and the
RAS program ``Mathematical Control Theory''. J.\ E.\ thanks the EU (QESSENCE, COMPAS, MINOS) and the EURYI
for support.

\appendix
\section{An important identity} \label{appA}
In this Appendix, we prove the important identity~(\ref{ideimpo})
 and  the inequality~(\ref{ineimpo}), by the following more general lemma.
\begin{lemma} {Let ${A}, {B} \in {Gl} ({m},
\rr)$ be  ${m} \times {m}$ real matrices with ${B}$ being skew-symmetric, which satisfy the inequality}
\begin{eqnarray}
{A} \geqslant i {B} \;. \label{dies}
\end{eqnarray}
{Then given ${A}^{\ominus1}$ the MP inverse~\cite{nota1} of ${A}$,
 the following identity holds}
{\em \begin{eqnarray} \label{ide}
2 \; \rank[{A}-i {B}] = \rank[{A}] + \rank[{A}- {B} {A}^{{\ominus1}}\,  {B}^T ]\; . \;
\end{eqnarray}}
Furthermore the following inequality applies
{\em \begin{eqnarray} \label{ine}
\rank[B]\geqslant  \rank[{A}] - \rank[{A}- {B} {A}^{{\ominus1}}\,  {B}^T ]\; . \;
\end{eqnarray}}
\end{lemma}

{\em Proof:} Let us start by reviewing some general properties of $A$ and $B$.
Because of Eq.~(\ref{dies}) the matrix ${A}$ must be positive semi-definite, and its support must contain the support of ${B}$. Consequently indicating with ${a}=\rank[{A}]$ and ${b}=\rank[{B}]$ the ranks of the two matrices,
we must have ${a} \geqslant  {b}$ with $b$ even.
Furthermore, defining  $\Pi\in  {Gl} ({m},\rr)$ to be the projector on the support of ${A}$, it will commute  with ${A}$ and ${B}$ and hence satisfy the following identity
\begin{eqnarray}
\Pi\;  {A}  = A \;\Pi = A \;, \qquad \qquad \Pi  \; {B} = B \; \Pi = {B}\;.
\end{eqnarray}
Consider then the
invertible matrix
\begin{eqnarray} \label{invert}
\bar{{A}}:= {A} + (\openone_{{m}} -\Pi)\; .
\end{eqnarray}
The MP inverse~\cite{HJ}  of ${A}$ is defined by
\begin{eqnarray}
{A}^{\ominus1}:= \Pi \bar{{A}}^{-1} \Pi \;.\end{eqnarray}
 To prove the validity of Eq.~(\ref{ide}) we note that it is possible to identify a congruent transformation
${A} \mapsto {A}' =C {A} C^T$,  ${B} \mapsto {B}' =C {B} C^T$,  with
$C \in {Gl} ({m},
\rr)$ invertible such that,
\begin{eqnarray} \label{y1111}
{A}' = {\footnotesize \left[
    \begin{array}{c|c}
 \openone_{a} & 0 \\ \hline
    0 & {0}
    \end{array}
    \right]  \begin{array}{l}
    \} \, a \\
    \} \,  {m} -a \;,
      \end{array}
    }
 \end{eqnarray}
and
 \begin{eqnarray} \label{sigma2222}
 {B}' = {\footnotesize \left[
    \begin{array}{c|c}
    \begin{array}{c|c}
  \begin{array}{c|c}
  0 & \mu \\ \hline
  -\mu  & 0
  \end{array} & 0 \\ \hline
  0 & 0
    \end{array}
    & 0 \\
\hline
0 & 0
  \end{array}
    \right]
    \begin{array}{l}
    \} \, b/2  \\ \} \, b/2 \\ \} \, a-b \\
    \} \, {m} -a \,,
      \end{array}
    \;,}
 \end{eqnarray}
 with $\mu = \mbox{diag}(\mu_1, \mu_2, \cdots, \mu_{b/2})$ being the $b/2\times b/2$ diagonal matrix formed by the strictly positive eigenvalues of $|{B}'|$ (by construction they satisfy
$1 \geqslant \mu_j \geqslant 0$).  The matrix $C$ can be explicitly constructed as follows.  First we identify the orthogonal matrix $O \in {Gl} ({m},
\rr)$ which diagonalizes ${A}$ and  $\Pi$ puts them in the following block forms:
\begin{eqnarray} \label{{A}prime1}
O{A}O^T= {\footnotesize \left[
    \begin{array}{c|c}
 {A}'' & 0 \\ \hline
    0 & 0
    \end{array}
    \right]  \begin{array}{l}
    \} \, a \\
    \} \,  {m} -a
      \end{array}
    } \; ,
 \end{eqnarray}
 \begin{eqnarray}
O \; \Pi \; O^T =  {\footnotesize \left[
    \begin{array}{c|c}
 \openone_{a} & 0 \\ \hline
    0 & 0
    \end{array}
    \right]  \begin{array}{l}
    \} \, a \\
    \} \,  {m} -a
      \end{array}} \; ,
    \end{eqnarray}
 with ${A}''\in  {Gl} (a,
\rr)$ being a $a\times a$ positive definite diagonal matrix.  Then we construct
the invertible matrix $K\in {Gl} ({m},
\rr)$ defined as
\begin{eqnarray} \label{KKK1}
K = {\footnotesize \left[
    \begin{array}{c|c}
 {A}''^{-1/2}  & 0 \\ \hline
    0 & \openone_{{m}-a}
    \end{array}
    \right]  \begin{array}{l}
    \} \, a \\
    \} \,  {m} -a
      \end{array}
    } \; ,
 \end{eqnarray}
(notice that the matrix ${A}''^{-1/2}\in {Gl} (a,
\rr)$  is well defined since ${A}''\in {Gl} (a,
\rr)$ is invertible). Finally, we take $O'\in {Gl} (a,
\rr)$ to be an orthogonal $a\times a$ matrix and define $C$ as follows
\begin{eqnarray} \label{Cprime}
C=
 {\footnotesize
 \left[
    \begin{array}{c|c}
 O'   & 0 \\ \hline
    0 & \openone_{{m}-{a}}
    \end{array}
    \right] K O }=
 {\footnotesize
 \left[
    \begin{array}{c|c}
 O' {{A}''}^{-1/2}  & 0 \\ \hline
    0 & \openone_{{m}-{a}}
    \end{array}
    \right]  O } \; \; . \ \ \ \  \;
 \end{eqnarray}
 By construction we have that for all the choices of $O'$ the resulting matrix is invertible and Eq.\ (\ref{y1111}) is satisfied.
 Vice versa, Eq.~(\ref{sigma2222}) can be satisfied by noticing that, since the support of ${B}$ is included into the support of ${A}$, we must have
 \begin{eqnarray} \label{{A}prime}
K O\; {B}\;  O^T K^T = {\footnotesize \left[
    \begin{array}{c|c}
 {B}'' & 0 \\ \hline
    0 & 0
    \end{array}
    \right]
     \begin{array}{l}
    \} \, {a} \\
    \} \,  {m} -{a}
      \end{array}
    } \; ,
 \end{eqnarray}
 with ${B}''\in {Gl} ({a},
\rr)$ skew-symmetric and having the same rank as ${B}$. By using a theorem from linear algebra
one can then find an orthhogonal
$O' \in {Gl} ({a},
\rr)$ such that
\begin{eqnarray}
 O'\; {B}''\;  O'^T =  {\footnotesize \left[
    \begin{array}{c|c}
  \begin{array}{c|c}
  0 & \mu \\ \hline
  -\mu  & 0
  \end{array} & 0 \\ \hline
  0 & 0
    \end{array} \right]}\;,
\end{eqnarray}
with $\mu$  being a positive diagonal matrix of dimension equal to the rank of ${B}''$  (the elements $\pm i \mu_j$ are its not null eigenvalues).
Using such an $O'$ in order to build $C$ as in Eq.~(\ref{Cprime})
we can then satisfy Eq.~(\ref{sigma2222}).

Now we notice that, since any congruent transformation preserves the rank of a matrix, the following identity holds:
\begin{eqnarray} \label{fuffa}
&&\rank[ {A} - i{B}] = \rank[ C( {A} - i {B}) C^T] \nonumber \\
&&=\rank{\footnotesize \left[
    \begin{array}{c|c}
    \begin{array}{c|c}
  \begin{array}{c|c}
  \openone_{b/2} & -i \mu \\ \hline
  i \mu  & \openone_{b/2}
  \end{array} & 0 \\ \hline
  0 & \openone_{a-b}
    \end{array}
    & 0 \\
\hline
0 & 0
  \end{array}
    \right]  }
     =
 a- \#_1(\mu)\;, \nonumber
 \end{eqnarray}
 where $\#_1(\mu)$ counts the number of eigenvalues of the matrix $\mu$ which are equal to $1$.
  The last identity  follows from
  counting the non-zero eigenvalue of the matrix on the left-hand-side of the second line. This can be easily done by observing that
   its spectrum contains ${m}-a$ explicit zeros (these are
 the terms in the zero block diagonal term), $a-b$ ones (these are the ones on the diagonals of the first block) and
  $1\pm \mu_j$  with $\mu_j \in [1,0]$ being the eigenvalues of $\mu$. Consequently,
  the non-zero eigenvalues are obtained by subtracting from $k$ (rank of the first block) the
  number $\#_1(\mu)$ of eigenvalues of $\mu$ which are equal to $1$. To compute
  the latter quantity we note that
 \begin{eqnarray}
{B}' {B}'^T
 = {\footnotesize \left[
    \begin{array}{c|c}
  \begin{array}{c|c}
  \begin{array}{c|c}
 \mu^2 & 0 \\ \hline
  0 & \mu^2
  \end{array} & 0  \\ \hline
 0 &   0
  \end{array}
 &  0
  \\ \hline
  0 & 0
    \end{array}
    \right]
    \begin{array}{l}
    \} \, b \\  \} \, b \\  \} \, a-b \\
    \} \, {m} -a \,,
      \end{array}
    \;,}
 \end{eqnarray}
 which yields
 \begin{eqnarray}  \label{hh1}
\rank[ \openone_{{m}} - {B}'{B}'^T] =  {m} - 2 \; \#_1(\mu) \;.
 \end{eqnarray}
Using the fact that $C$ is invertible, one has
  \begin{eqnarray}
&&\rank[ \openone_{{m}} - {B}'{B}'^T] = \rank[ \openone_{{m}} - C {B} C^T C {B}^T C^T] \nonumber \\&& \qquad \qquad =\label{important}
 \rank[ C^{-1} C^{-T}  -  {B} C^T C {B}^T ] \;.\nonumber
 \end{eqnarray}
 Since $O'$ and $O$ are orthogonal, we notice that $C^{-1} C^{-T}$ is composed of
 two terms that span orthogonal supports. Specifically we can rewrite it as
 \begin{eqnarray} &&
 C^{-1} C^{-T}  = O^T K O = O^T
 \;{\footnotesize \left[ \begin{array}{c|c}
 {A}'' & 0 \\ \hline
    0 & \openone_{{m}-a}
    \end{array}
    \right]
    } \;
     O \nonumber \\ &&= {A} + O^T  \;{\footnotesize \left[ \begin{array}{c|c}
 0 & 0 \\ \hline
    0 & \openone_{{m}-a}
    \end{array}
    \right]
    } O  = {A} + (\openone_{{m}} -\Pi) = \bar{{A}} \;,\nonumber \end{eqnarray}
    where Eqs.~(\ref{{A}prime1}) and (\ref{KKK1}) have been used.
    Similarly, ${B} C^T C {B}^T$ is only supported on the support of ${A}$. Indeed, we have
    \begin{eqnarray}
    \nonumber
    &{B} C^T C {B}^T =   (\Pi\;  {B} \; \Pi) \;  [C^{-1} C^{-T}]^{-1} \; (\Pi \; {B}^T \; \Pi)& \\
    &= (\Pi\;  {B} \; \Pi) \;  \bar{{A}}^{-1} \; (\Pi \; {B}^T \; \Pi) = (\Pi\;  {B} \; \Pi^2)  \;  \bar{{A}}^{-1} \;   (\Pi^2 \; {B}^T \; \Pi) & \nonumber \\
    &=(\Pi\;  {B} \; \Pi)  (\Pi\; \bar{{A}}^{-1}\; \Pi)  (\Pi \; {B}^T \; \Pi)  & \nonumber \\
    &= (\Pi\;  {B} \; \Pi)  {A}^{{\ominus1}}\,  (\Pi \; {B}^T \; \Pi) =  {B} \;   {A}^{{\ominus1}}\,   {B}^T \;.\label{uu1}&
        \end{eqnarray}
 Using these identities, we can then rewrite Eq.~(\ref{important}) as
 \begin{eqnarray} \nonumber
&&\rank[ \openone_{{m}} - {B}'{B}'^T]
 =\rank[  \bar{{A}}- {B} {A}^{{\ominus1}}\,  {B}^T   ]  \nonumber \\
 &&=\rank[\openone_{{m}} -\Pi] +
 \rank[  {{A}}- {B} {A}^{{\ominus1}}\,  {B}^T   ]  \nonumber \\
    && = {m} - a +  \rank[{A}-{B} {A}^{{\ominus1}}\,  {B}^T  ]\;.
    \label{b18}
 \end{eqnarray}
Thanks to Eq.~(\ref{hh1}) the above identity finally yields
 \begin{eqnarray}\label{ggfi}
  \#_1(\mu) =\frac{ \rank[{A}] - \rank[{A}- {B} {A}^{{\ominus1}}\,  {B}^T ]}{2}\;,
  \end{eqnarray}
which gives  Eq.~(\ref{ide}) when inserted into Eq.~(\ref{hh1}).
The inequality (\ref{ine}) can finally be proven by noticing that because of the invertibility of $C$, one has
$\rank[B] =\rank[B'] = b$ which, by construction, is larger than $2 \#_1(\mu)$. The result then
follows simply by applying Eq.~(\ref{ggfi}). $\blacksquare$
\section{Gaussian purifications}\label{ap:B}
Here, we emphasize the minimal number of ancillary modes which are necessary to construct a Gaussian purification (i.e.,
a purification which is joint pure Gaussian state of the system and of the ancillary modes)
 of a generic multi-mode Gaussian state $\hat{\rho}$. Of course,
the Gaussian requirement on the purification is fundamental for our purposes: Since any number of modes can always be
embedded in a single one, by dropping it the minimal number of modes is always smaller than or equal to one.

\subsection{Minimal Gaussian purifications of Gaussian mixed states} \label{firstsub}
Let $\gamma \in {Gl} ({2n},
\rr)$ the covariance matrix of a  Gaussian state $\hat{\rho}$ of a system $A$ formed by $n$ bosonic modes.
We know that it must satisfy the following inequality
\begin{eqnarray}\label{HB}
	\gamma \geqslant i \sigma_{2n}\;,
\end{eqnarray}
with $\sigma_{2n} \in {Gl} ({2n},
\rr)$ being the skew-symmetric matrix in Eq.~(\ref{sigma2n}) representing the symplectic form of the modes.
Thanks to the Williamson's theorem~\cite{WILLI} we know that there exists a symplectic transformation $S\in {Gl} ({2n},
\rr)$ which allows us to diagonalize $\gamma$ in the following form
\begin{eqnarray}\label{W}
\gamma \mapsto S \gamma S^T = {\footnotesize \left[
    \begin{array}{c|c}
D& 0 \\ \hline
 0 & D
    \end{array}
    \right]     \;,}
 \end{eqnarray}
 with $D\in {Gl} ({n},
\rr)$ being the diagonal matrix formed by the symplectic eigenvalues $D_j$  of $\gamma$ which satisfy the condition $D_j \geqslant 1$
as follows from Eq.\ (\ref{HB}). The values $\{D_j\}$ are the {\it symplectic eigenvalues} of $\gamma$ \cite{HOLEVOBOOK,Review}, so
the positive square roots of the eigenvalues of the matrix $- \sigma_{2n} \gamma\sigma_{2n} \gamma\in Gl(2n,\rr)$.
The transformation $\gamma\mapsto
S \gamma S^T$ corresponds to appling a Gaussian unitary to the state which transforms it into a product state of the $n$ modes, in fact
a product of Gibbs states of unit harmonic oscillators. Hence, it does not restrict generality to assume that $\gamma$ is of the
form of the rhs.\ of Eq.\ (\ref{W}) in the first place.

Now, let $\Gamma$ be the covariance matrix of
the minimum purification of $\gamma$, viewed as being defined on a bi-partite system labeled $A$ --- the original system
--- and $B$. Since the spectrum of the reduced state with respect to $B$ is identical to the spectrum of the reduced state
of $A$, also the symplectic spectra of the two reductions are the same. Hence, it does not restrict generality to take $\Gamma$ to be
of the form
\begin{equation}
	\Gamma = {\footnotesize \left[
	\begin{array}{c|c}
    		\begin{array}{c|c}
		D& 0 \\ \hline
 		0 & D
   		 \end{array}
    	&
		C\\ \hline
	C^T &
		\begin{array}{c|c}
		D& 0 \\ \hline
 		0 & D
   		 \end{array}
	\end{array}
	\right]   }  ,	
\end{equation}
with suitable $C\in Gl(2n,\rr)$ such that the symplectic spectrum of $\Gamma$ consists of $1$ only, with respect to the symplectic
form in the convention as in Eq.\ (\ref{NS}).
Now, by taking
\begin{equation}
	C =  {\footnotesize \left[ \begin{array}{c|c}
		0& \eta \\ \hline
 		\eta & 0
   		 \end{array}\right]}\;,
\end{equation}
with $\eta= \mbox{diag}(f(D_1),\cdots , f(D_n))$, one clearly arrives at the covariance matrix of a valid
purification. This purification essentially involves as many modes as there are symplectic eigenvalues
different from $1$ --- those modes associated with unit symplectic eigenvalues correspond to pure Gaussian states
already. Denoting the number of unit values in $D$ by $\#_1(D)$, this purification hence involves $n-\#_1(D)$ many modes.
Invoking the definition of the symplectic spectrum, one finds that
\begin{equation}
	\#_1(D)= n -\rank[\gamma -\sigma_{2n} \gamma^{-1} \sigma_{2n}^T] /2\;.
\end{equation}
It is also easy to see, however, that no purification can involve fewer modes than that. Consequently we have
\begin{equation}
	q_{\min}= n-\#_1(D).
\end{equation}
The covariance matrix of the reduced Gaussian state of the purification with respect to $B$ is necessarily given by the
rhs.\ of Eq.\ (\ref{W}), up to local symplectic transformations $S\in Sp(2n,\rr)$.
Hence, any Gaussian purification
must involve at least involve $n-\#_1(D)$ modes, as so many symplectic eigenvalues are different from $1$. Needless to say,
if one gives up the property of requiring a Gaussian purification, one can always embed the purification
in a single mode, if the state is mixed, while no additional mode being required if the state is already pure.


\begin{thebibliography}{99}
\bibitem{footnote}
	Namely, if ${{\hat{\rho}}}_E=|\psi_E\rangle\langle\psi_E |$, then Eq.\ (\ref{first})
	is equivalent to the Stinespring representation $\Phi^*(\hat X)=\hat V^\dagger (\hat X\otimes {\cal I}_E)\hat V$
	for the dual channel in the Heisenberg picture, where the isometry 	
	$\hat V$ is given by $\hat V=\hat{U}_{A,E}|\psi_E\rangle$.
\bibitem{NOTA}
	Note that for a  positive semi-definite matrix $M$ we have
	$\mbox{ker}[M]=\left\{  w\in \mathbb{C}^{d}:w^{\dag}Mw =0\right\}$.
\bibitem{nota1}
	Given $\Pi$ the projector on the support of $Y$ and defined the
	invertible matrix $\bar{Y}:= Y + (\openone -\Pi)$, the MP-inverse of $Y$  is defined as
	$Y^{\ominus1}:= \Pi \bar{Y}^{-1} \Pi$ (see also Ref.\ \cite{HJ}).
\end{thebibliography}
\end{document}